\documentclass{elsart}

\begin{document}

\begin{frontmatter}
\title{ Test of final state approximations using
threshold $pp \rightarrow pp\pi^{0}$ }

\author{J. A. Niskanen}

\address{Department of Physics, P. O. Box 9 \\
FIN-00014 University of Helsinki, Finland\thanksref{email}}
\thanks[email]{Email: jouni.niskanen@helsinki.fi}
 
 
\begin{abstract}
The Watson-Migdal approximation scheme to take into
account final state interactions is shown to give the
actual threshold momentum dependence of the reaction 
$pp \rightarrow pp\pi^{0}$. However, by an explicit
plane wave replacement of the final state wave function
it is stressed that not too much physical significance
should be given to the proportionality coefficient 
extracted using this procedure. The plane wave
approximation is not physically reliable even after 
introducing  the Watson-Migdal or a more sophisticated
final state interaction factor, since direct production
(impulse term) is missed. Also with short range interactions
there can be discrepancies of a factor two.


\end{abstract}

\end{frontmatter}

During the 90's meson production at threshold has raised
much experimental activity and interest with  cooler
facilities  producing data with unprecedented accuracy 
and energy resolution in a region where only a single
partial wave amplitude should contribute. One of the
exciting results is the threshold cross section of the
reaction $pp \rightarrow pp\pi^{0}$  \cite{meyer,bondar},
others involve e.g. $\eta$ meson production \cite{etaexpt}.
This activity has also revived theoretical interest with
many mechanisms proposed for each reaction. Discussion
of these mechanisms is not the main purpose of this paper.

Most theoretical approaches are based on the established
and apparently in this case numerically well justified 
DWBA including various irreducible production mechanisms
between initial and final states calculated for realistic
interactions. There is, however, one class of works that use 
strong assumptions about the possibility to approximate
the $NN$ wave functions either by using apparently a plane
wave Born approximation or including just the low energy
final state interaction (FSI) approximately.
In the latter it is assumed that the effect of the FSI can
be factorized from the transition matrix elements and
be included after the matrix elements have been calculated 
using simple analytic wave functions, notably plane waves.
Among such approaches - while having the value of introducing
the mechanisms at a fundamentally more basic level
as relativistically covariant or by chiral
perturbation theory - are Refs. \cite{engel,bernard}
on $\pi^0$ production and Refs. \cite{moalem} on $\eta$
production. The aim of this paper is to test how well such 
approximations reflect the numerical (and partly physical)
reality by an explicit calculation of $pp \rightarrow pp\pi^{0}$.\\

At low energy or threshold scattering and reactions an often 
used approximation to take into account the final state 
interaction is the Watson-Migdal procedure \cite{watson}.
This consists of taking first the asymptotic scattering
(final) state wave function with the relative $NN$ momentum
$p_f$
\begin{equation}
{u_f( r)\over p_f r} = 
{  \sin (p_f r+\delta)\, e^{i\delta} \over p_f r}
\end{equation}
(only the nucleon $S$-wave is considered here) and then
extending this asymptotic form to the range of nuclear
forces (where $p_f r \approx 0$; $r$ cancels against the volume
element in integration)
\begin{equation}
{u_f(r)\over p_f} \rightarrow 
{\sin\delta\, e^{i\delta} \over p_f}
= {1 \over p_f \cot\delta - ip_f} .
\end{equation}
Here it is usual to make a further approximation in terms
of the scattering length by replacing
\begin{equation}
p_f \cot\delta \approx -\, {1\over a},
\end{equation}
where $a$ is the large scattering length. So in $S$ waves
one expects the threshold behaviour of the cross section to be
\begin{equation}
\sigma \sim {1\over p_f^2 \, (\cot ^2\delta + 1) } \times p_f 
\sim {a^2\over 1 + (p_f a)^2 } \times p_f ,
\end{equation}
where the $p_f$ in the numerator is the momentum dependence
of the phase space. If the infinite ranged Coulomb force
is present in the final state the cross section should behave
like
\begin{equation}
\sigma = {1\over C_0^2 }
 {1\over p_f^2 (\cot ^2\delta + 1) } \times {\rm phase\;
 space}\times{\rm const.}
\equiv F_{FSI}\times {\rm phase\; space}\times{\rm const.}\, , 
\end{equation}
where $\delta$ is now the Coulomb-strong phase shift and
$C_0^2$ is the Coulomb penetration factor
\begin{equation}
C_0^2 = {2\pi\eta \over e^{2\pi\eta} - 1 },\qquad
\eta = {e^2 M_p\over 2\hbar^2 p_f} \, .
\end{equation}

The remaining interaction matrix element is assumed to
be relatively constant as compared to the fast varying
$F_{FSI}$. However, it is easy to erroneously imply
some direct specific meaning(s) for the constant extracted
in this way from experiments. There is nothing in Ref.
\cite{watson} which would relate the actual reaction
matrix element to this constant, since the radial
dependence obtained from the asymptotic form
is meaningless at short distances. In particular this 
constant is {\it not} to be interpreted as the Born 
approximation to the reaction matrix element
and this interpretation is not advocated in the
original work. It only reflects the
general final momentum or energy dependence embedded
in $F_{FSI}$ for small momenta $p_f$. \\

To exemplify this in a specific reaction with the strongly
attractive nearly bound $^1S_0$ two-nucleon final state
I discuss very explicitly the reaction $pp \rightarrow pp\pi^{0}$
close to threshold. This has both complications:
the strong interaction and the Coulomb interaction in
the final state. The energy discussed is 290.7 MeV
corresponding to $p_f({\rm max}) = 0.35\, {\rm fm}^{-1}$
and is about 5 MeV above threshold in the c.m.s..
The formalism has been given elsewhere \cite{pppppi}
and will not be repeated here as not very essential for
the present argument. Suffice it to say that the pion
is produced primarily by the Galilean invariant $\pi NN$
vertex for each nucleon obtained from the pseudovector 
pion coupling 
\begin{equation}
 H_{\pi NN} = {f \over m_{\pi^+}}\ {\vec\sigma} \cdot 
\left\{  {\vec q}\  {\vec\tau} \cdot {\vec\phi} - 
 {\omega_q \over 2M} 
\left[ {\vec p} \ {\vec\tau}\cdot {\vec \phi} + {\vec\tau} \cdot 
{\vec\phi}\ {\vec p} \right] \right\} .
\end{equation}
Also it is necessary to include at least pion $s$-wave
rescattering from the second nucleon and some additional
mechanism, such as a heavy meson exchange with a 
nucleon z-diagram in the production vertex to fit the
data (without this, direct production is a gross underestimate 
as compared to data). In reactions where the $NN$ isospin changes
this $s$-wave rescattering even dominates at threshold, but
in the present case it is smaller than the direct axial charge
term. To avoid unnecessary complications
the treatment here considers only nucleons, no $\Delta$
components.

The conventional mechanisms lead to the exact transition 
amplitude at the two-nucleon level  
\begin{eqnarray} 
< ^1\! S_0 | H_{\rm prod} |^3P_0 >& = &\displaystyle{-8\pi  
\over p_f p_i\sqrt{\omega_q}}
{f\over m_{\pi^+}}
\left\{ (1-{\omega_q\over 2M})\, q\, \int u_f^* (r) j_1 (qr/2)u_i (r) dr
\right.\cr
&+ &{\omega_q\over M} \int u_f'^{*} (r) j_0 (qr/2)u_i (r)  dr \cr
&+ &{\lambda_1\over m_{\pi^+} } \left[ (2-{\omega_q\over 2M}) 
\int u_f^* (r) f'(r)j_0  ({qr\over 2})u_i (r) dr\right.\cr
&- &\left.\left. {\omega_q\over M} \int u_f'^{*}(r)f(r) 
j_0({qr\over 2})u_i(r) dr
\right]\right\}.
\end{eqnarray}
Here the first term arises from the standard $p$-wave $\pi N$ 
coupling proportional to the 
pion momentum $\vec\sigma\cdot\vec{q}$ and should be small 
at threshold due to the presence of the pion wave function
$j_1 (qr/2)$ in the integrand and the factor $q$. 
The Galilean invariance term
does not have this suppression, just the factor $\omega_q/M
< 1$. The last two terms involve 
$s$-wave pion rescattering off the second nucleon with the 
isospin symmetric  amplitude $\lambda_1 .$ The Yukawa function 
$f(r)$ and its derivative $f'(r)$  arise 
from the propagator of the intermediate pion and have a range 
rather similar to OPE \cite{koltun}.  
The wave function derivative in the present normalization 
(see Eq. (1)) is $u'(r)=r\,{\rm d} (u(r)/r) /{\rm d} r$. 
One may note that the latter terms involve an "interaction" of 
finite range, while the first two overlaps have an infinite 
range.  It is therefore of particular interest to see how 
well short range expansion of the final state wave function, 
the basis of the Watson-Migdal method, fares here as 
compared with an exact calculation. Of course, correlations 
are necessary for the reaction to happen at all, since
an on-shell nucleon cannot emit a real pion. 
So, actually the reaction does have a finite range, though
the integrals to be evaluated extend to infinity and no
convergence factor is present. This expectation will be
 borne out in
numerical results, where little difference is found between 
the energy dependences of the first and last terms.

The above integrals can be calculated numerically 
\cite{pppppi}.  The dominant terms near threshold 
are the Galilean invariance term (the second) and the 
rescattering terms. When the pion has enough energy,
the first term becomes comparable (in fact, physically
it will need also the inclusion of the $\Delta$). 
For the present study of the effect of the final state 
interaction, the dependence on the relative energy 
(or momentum $p_f$) of the final state nucleons is 
particularly relevant.  
So special care with also the long ranged Coulomb force 
is important, since it has large influence at low energies. 
In all cases presented here the initial state is the same
distorted $^3P_0$ wave function calculated from the Reid
soft core potential \cite{reid}. This potential is
also used for the final state in the "exact" case. 

\begin{figure}
\vspace{3cm}
\caption{
The ratio of the squared matrix elements and the final state
interaction factor $F_{FSI}$ defined in Eq. (5). Dotted curve:
direct axial current contribution (nongalilean) multiplied by 10; 
Dashed: full direct Galilean production operator $H_{\pi NN}$ used; 
Solid: full result including also pion $s$-wave rescattering;
Dash-dot: full result with added HME to fit the data at threshold
(divided by 10).
}
\end{figure}

Fig. 1 shows the ratio of the squared transition matrix
element (calculated using the exact wave function) to the
final state interaction factor $F_{FSI}$ as a function of 
the final state relative $pp$ momentum with the inclusion 
of different mechanisms. The dotted curve shows the small
effect of the direct axial current coupling (nongalilean
first term, multiplied by 10) 
$f/m_{\pi^+}\,\vec\sigma\cdot\vec q$, where
$\vec q$ is the pion momentum. The dashed curve includes
also the much more important axial charge coupling 
(Galilean part) $-f/m_{\pi^+}\,\omega_q\,\vec\sigma\cdot
(\vec p + \vec p')/2M$. In the solid curve also pion
$s$-wave rescattering is added to these impulse terms 
- in several pionic reactions
this is sufficient and satisfactory. By far, most of this
contributes through the nongalilean part. The minute
direct nongalilean result has a qualitatively different
energy dependence, since it has to vanish, when $q
\rightarrow 0$, i.e. when $p_f \rightarrow p_{f,max}$.

However, this reaction needs some additional mechanism 
\cite{meyer} and in the dash-dot curve a phenomenological
heavy meson exchange is added to reproduce the experimental 
cross section \cite{lee}. It can be
seen that in this momentum range, apart from the FSI
factor, the momentum dependence is moderate for all
mechanisms and the validity of the Watson-Migdal procedure 
is confirmed. It may also be noted that even the "long-ranged" 
galilean impulse term does have the same momentum 
dependence as those with an explicit short-range interaction.
This is, of course, due to the fact that the correlations 
necessary for the reaction are generated by meson exchanges.
If the Coulomb interaction is removed, the results are 
nearly the same: the ratios increase by 10-20\% and the
energy dependence is marginally weaker.

As another step we try to interpret the matrix element as 
arising simply from a use of a plane wave for the final state,
trying to take the FSI into account just by the factor $F_{FSI}$.
Fig. 2a shows the behaviour of the squared transition 
matrix for the direct (impulse) contribution as
a function of the relative final two-nucleon momentum $p_f$. 
The solid curve employs the exact final state
wave function (likewise in the other figures).
To make a "Born  approximation" with respect to the final state
one replaces $u_f(r)/(p_fr)$ in Eq. (8) by $j_0(p_fr)=
\sin(p_fr)/(p_fr)$. The resulting negligibly small 
approximate results are not shown in the figure.

\begin{figure}
\vspace{3cm}
\caption{
Momentum dependence of the squared transition matrix. 
Curves: Solid) exact final state; 
Dashed) final state approximated by a plane wave and 
multiplied by  $F_{FSI}$ of Eq. (5) (fm$^9$);
Dotted) exact $s$-wave rescattering without the impulse term;
Dash-dot) the FSI factor from Eq. (\protect\ref{jost}) used.
Windows: a) direct ("impulse") terms only, b) pion $s$-wave
rescattering included, c) also phenomenological heavy meson
exchange added, d) only HME. 
In a (direct terms) the approximate results are negligible 
(smaller by 4--6 orders of magnitude). }
\end{figure}

In fact, both approximate impulse amplitudes are much smaller 
than the corresponding exact ones, the nongalilean one by
two orders of magnitude, the galilean one by three orders.
This smallness is apparently due to the overall orthogonality
of the three relevant Bessel functions \cite{jackson} and
to the fact that the $^3P_0$ initial wave function does
not deviate much from the plane wave (at 290.7 MeV the 
calculated phase shift is $-9.5^\circ$). It may be noted
that also the derivative of the final $S$-wave as defined
above is a Bessel function. Making the $NN$ potential
stronger increased these amplitudes; making it zero further
reduced the direct amplitudes by another two orders of magnitude
acting as a check on numerics of these infinite-ranged
oscillatory integrals. 

The strong $S$-wave final state
interaction removes this orthogonality making the amplitude
sizable. These correlations influence particularly much on
the derivative, making the matrix element of the
galilean term physically important.
Also inclusion of an explicit interaction such as pion
rescattering removes the orthogonality in the integrand.
Then apparently at threshold the approximate integral 
is relatively insensitive to $p_f$, since $j_0(p_fr)
\approx 1$ for small $r$.

With a finite ranged interaction the energy dependences of
the approximate and exact results are qualitatively
similar (dashed curves vs. solid in Figs. 2b-d),
once the FSI factor is applied, as would be obvious from
the constancy of the approximate integral.
 However, physically in
the exact result it is necessary to take into account the
significant galilean impulse term. In the  galilean 
case (Fig. 2b) the impulse term is constructive with
rescattering and the exact cross section is larger than the 
approximate FSI result, while in the nongalilean result the 
situation would be reversed. Even the FSI factor cannot
reproduce the full height of the exact result, even though
the approximate rescattering is more than twice as large as 
the exact (dotted curve in 2a). This shows clearly the 
physical importance of the galilean impulse term. 

Finally in Fig. 2c the overestimation of short ranged
mechanisms by the approximate factor (shown explicitly in 2d)
reproduces the exact result (with a slight overestimate).
However, this good agreement is achieved by overestimation of
the short-ranged interactions and omission of direct production.

Now it is important to note that, since also the approximate
final wave function includes the $1/p_f$, it is {\it not} 
consistent to apply the FSI factor defined by Eqs. (2--5)
where the same $1/p_f$ appears -- the result would  not 
be even dimensionally correct. If this double counting is
avoided, the resulting energy dependence clearly would 
disagree with the exact result - the dashed curves should
be multiplied by $p_f^2$. This exercise is done, because
a more sophisticated treatment is not numerically very 
different (if $p_f$ is given in fm$^{-1}$) and because it
is possible that the simpler method may have been applied
some times in the past. Also, it was seen that the energy
dependence comes out correctly.

An enhancement factor of the amplitude having a correct 
high-energy limit is based on properties of the Jost function 
and presented in Ref. \cite{gw} for a short-ranged interaction
applying the effective range expansion to order $p_f^2$
\begin{equation}
\frac 1{{\bf f}(-p_f)} = \frac{(p_f^2+\alpha^2)r_0/2}
{1/a + r_0p_f^2/2 - ip_f}  \label{jost}
\end{equation}
with 
\begin{equation}
\alpha =  (1 + \sqrt{1 + 2r_0/a\,}\, )/r_0 .
\label{alpha}
\end{equation}
Except for the numerator, to this order this is equivalent to
the simpler form and is dimensionless. Also, if the unit of
length is the femtometer, the numerator is indeed of the order
of unity for small $p_f$. In Ref. \cite{engel} this is used
with a Coulomb modification on the scattering length $a = 7.8243$ 
fm and effective range $r_0 = 2.7058$ fm given now as \cite{shyam}
\begin{eqnarray}
& 1/a_c & =  [1/a - 2 p_f \eta h(\eta)]/C_0^2 \\
& r_{0c} & =  r_0/C_0^2 
\end{eqnarray}
with the function $h(\eta)$ given on p. 263 of Ref. \cite{gw}.
However, since $a_c$ and $r_{0c}$ are not constant with
$p_f$, it may not be clear that the assumptions on which
Eq. (\ref{jost}) is based are valid in the presence of the
long-ranged Coulomb force. The numerical error arising from
this complication is likely quite small but not under control. 

Now using this improved final state interaction factor the 
dash-dot curves in Fig. 2 show similar good qualitative energy
dependence with the exact result as the dashed ones -- only 
slightly weaker. However,
although the normalization is much improved (as can be seen
from the dash-dot vs. dotted curve in 2b and dash-dot vs. solid
in 2d), still a mere
multiplicative factor cannot correct for the missing direct
production strength (the dash-dot vs solid curves in 2b and 2c).
Consequently, even 
after the inclusion of the heavy meson exchange the total 
results remain now as underestimates shown in Fig. 2c, 
while only this short range contribution is
presented in Fig. 2d to see the effect of the range alone.
In this case the approximation is a slightly larger 
overestimate, about a factor of three for the simple
factor. 

Above, the Coulomb interaction has been included in the 
final state interaction. It is also interesting to see how
the approximation fares without this complication. After all,
Eqs. (\ref{jost}-\ref{alpha}) are deduced for  the uncharged
effective range parametrization. One might consider this to
describe e.g. $nn \rightarrow nn\pi^0$ (hardly physically
measurable), $np \rightarrow nn\pi^+$ or $np \rightarrow np\pi^0$
reactions. However, to avoid superficial differences the
kinematics is kept the same relevant to $pp \rightarrow pp\pi^0$.
Even so, the scattering length and effective range are taken
to be the same experimental $np$ singlet parameters $a_{np} =
23.715$ fm and $r_{0np} = 2.73$ fm as in Ref. \cite{engel}.
Use of the 
Coulomb modified quantities above relevant to $pp$ scattering
would hardly be physically meaningful and would give an
incorrect view of the effect. 

\begin{figure}
\vspace{3cm}
\caption{As Fig. 2 but without the final state Coulomb force}
\end{figure}

The results are shown in Fig. 3
in the same way as in Fig. 2. It can be seen that now both 
approximations are close to each other, since the numerator
of Eq. (\ref{jost}) is (incidentally) close to unity, if the
momentum is given in fm$^{-1}$. Now both approximate 
methods give an overestimate by factor two for 
matrix elements involving potential ranges. The total 
result now agrees incidentally with the exact result but
due to the overestimate of HME and omission of direct 
production as earlier for the simpler procedure. Also, it
can be seen that for small values of $p_f$ the functional
dependence changes qualitatively. 

In summary, it was seen that the Watson-Migdal conjecture 
that at low momenta the reaction matrix momentum dependence
can be obtained from the asymptotic scattering wave function
(essentially the phase shift) is true in this reaction.
However, using this procedure in the other direction by
calculating the reaction matrix by plane wave functions and 
simply applying the final state interaction factor of the 
Watson-Migdal method or by the method given in Ref. \cite{gw}
is risky and may lead to physically incorrect conclusions 
about the reaction mechanisms. At its simplest, the energy 
dependent enhancement factor $F_{FSI}$ cannot be correctly 
used with the Born approximation potential matrix elements, 
whereas the more sophisticated procedure gives them reasonably
well. Both approximations overestimate short-ranged
mechanisms when the Coulomb force is not present in the final
state. 

In particular, in $pp \rightarrow pp\pi^0$ the multiplicative 
procedures miss the important direct production mechanism. In
principle, it is possible to incorporate most important meson
exchanges after the pion production vertex. However, this may 
lead to strong violation of unitarity and also obscures the 
role of the FSI. Inclusion of any FSI factor would then risk 
doubly counting parts of this interaction.\\

\begin{ack}
I thank Harry Lee for urging me to do this study.
This work was partly supported by the Academy of Finland.
\end{ack}

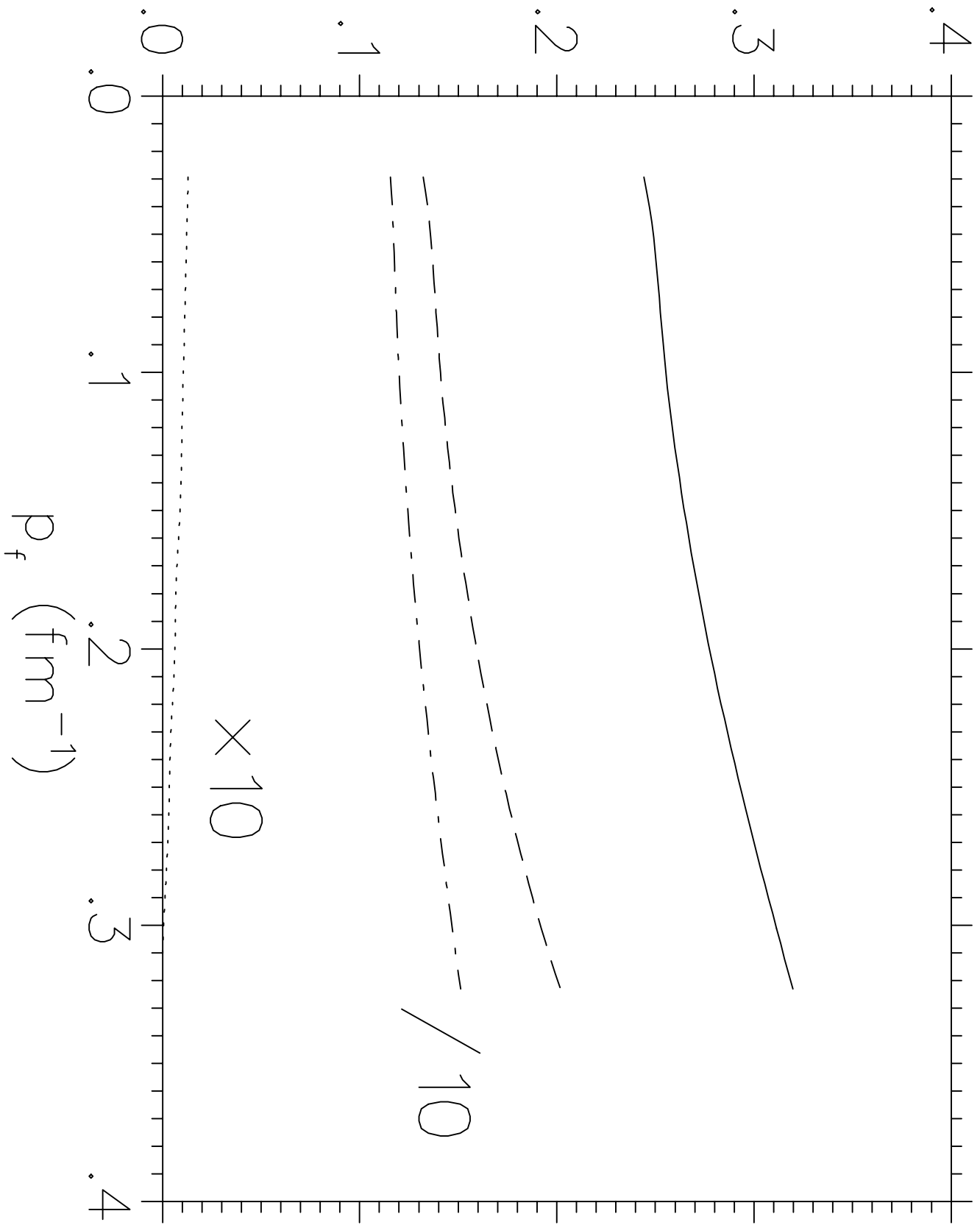
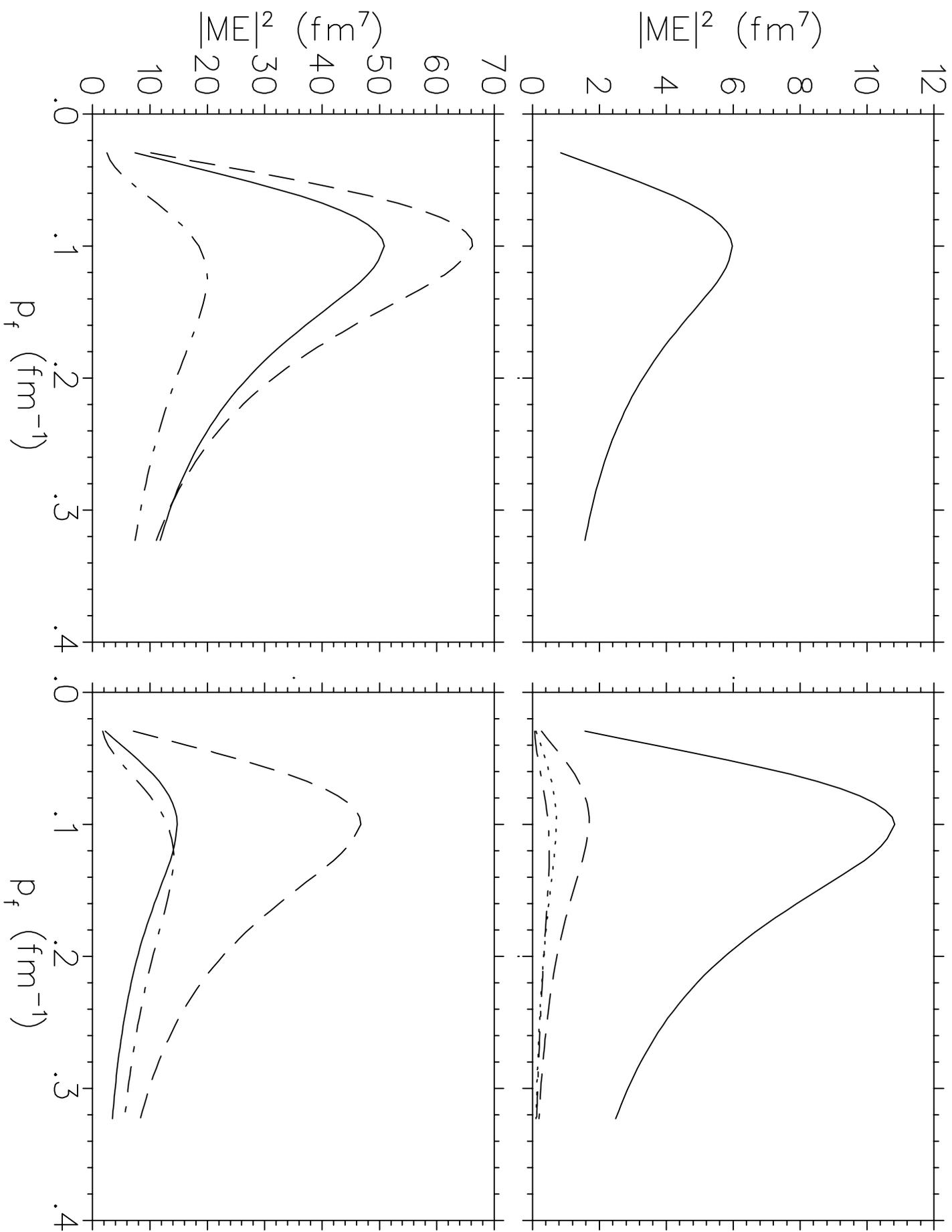
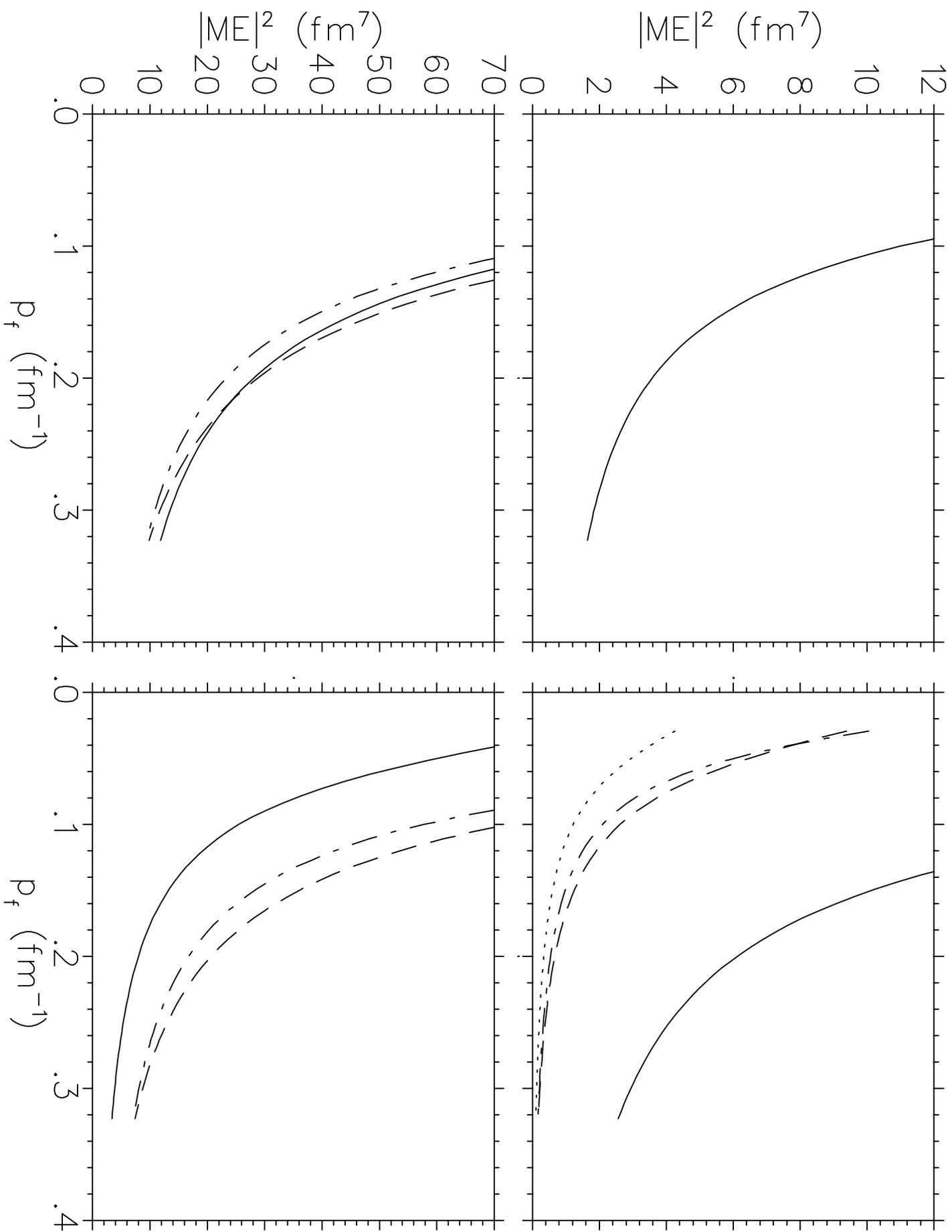
\end{document}